%Paper: hep-th/9403119
%From: Michael Douglas <mrd@mike.rutgers.edu>
%Date: Mon, 21 Mar 94 11:24:10 -0500

%
% a uuencoded figure is supplied in a separate file.
% the following is to incorporate it in the body of the text,
% using epsf.
\let\includefigures=\iftrue
%
% activate this if you don't have epsf.
%\let\includefigures=\iffalse
%
% the following is to use blackboard bold fonts --
\let\useblackboard=\iftrue
%
% activate this if you don't have them.
%\let\useblackboard=\iffalse
%
%
\input harvmac.tex
\includefigures
\message{If you do not have epsf.tex (to include figures),}
\message{change the option at the top of the tex file.}
\input epsf
%\epsfclipon
%\def\fig#1#2{\topinsert\epsffile{#1}\noindent{#2}\endinsert}
\def\fig#1#2{\midinsert\epsffile{#1}\noindent{#2}\endinsert}
\else
\def\fig#1#2{}
\fi
\def\Title#1#2{\rightline{#1}
\ifx\answ\bigans\nopagenumbers\pageno0\vskip1in%
\else%\special{papersize=11in,8.5in}%
\pageno1\vskip.8in\fi \centerline{\titlefont #2}\vskip .5in}

\ifx\answ\bigans\def\tcbreak#1{}\else\def\tcbreak#1{\cr&{#1}}\fi
\useblackboard
\message{If you do not have msbm (blackboard bold) fonts,}
\message{change the option at the top of the tex file.}
\font\blackboard=msbm10 scaled \magstep1
\font\blackboards=msbm7
\font\blackboardss=msbm5
\newfam\black
\textfont\black=\blackboard
\scriptfont\black=\blackboards
\scriptscriptfont\black=\blackboardss
\def\Bbb#1{{\fam\black\relax#1}}
\else
\def\Bbb{\bf}
\fi
% *************************************
%\draft
%
\def\SDiff{{\rm SDiff}~T^2}
\def\Cr{\hfill\break}
\def\comments#1{}
\def\BR{\Bbb{R}}
\def\BZ{\Bbb{Z}}
\def\BC{\Bbb{C}}

\def\p{\partial}
\def\half{{1\over 2}}
\def\Tr{{\rm Tr\ }}

\def\im{{\rm Im\hskip0.1em}}

\def\ket#1{|#1\rangle}
\def\vev#1{\langle{#1}\rangle}
\def\ppt{{\partial\over\partial t}}

\Title{\vbox{\baselineskip12pt
\hfill{\vbox{
%\hbox{others}\hfil}
\hbox{RU-94-29\hfil}}}}}
%Large $N$
{\vbox{\centerline{Chern-Simons-Witten Theory}
\medskip\centerline{as a Topological Fermi Liquid}}}
\centerline{Michael R. Douglas}
\smallskip
\centerline{Dept. of Physics and Astronomy}
\centerline{Rutgers University }
\centerline{\tt mrd@physics.rutgers.edu}
\bigskip
\noindent
We reinterpret $U(N)$ Chern-Simons-Witten theory
quantized on a torus as a free fermion system.
Its Hilbert space and some observables are simply related to those of group
quantum mechanics, even at finite $N$ and $k$.
Its large $N$ limit can be described using techniques developed for
matrix quantum mechanics and two-dimensional Yang-Mills theory.
We discuss the bosonization of this theory, which for YM$_2$ gave a precise
interpretation of Wilson loop operators in terms of string creation and
annihilation operators,
and examine its consequences for a string interpretation here.
The formalism seems entirely adequate for the leading large $N$ results and in
a sense can be thought of as a `classical string field theory'.
In considering subleading orders in $1/N$,
we identify some major differences between CSW and YM$_2$,
which must be dealt with to find a CSW gauge string interpretation.
Although these particular differences are probably not relevant for `QCD
string,' they do illustrate some of the issues there, and we comment on this.
We also propose an approach to dealing with large $N$ transitions.

\Date{March 20, 1994}
\nref\BIPZ{E. Brezin, C. Itzykson, G. Parisi and J.-B. Zuber, Comm.~Math.~Phys
59 (1978) 35.}
\nref\bosecone{D. J. Gross and I. Klebanov, Nucl.~Phys.~B352 (1990) 671;\Cr
A. M. Sengupta and S. R. Wadia, Int.~J.~Mod.~Phys.~A6 (1991) 1961.}
\nref\polchinski{J. Polchinski, Nucl.~Phys.~B362 (1991) 125.}
\nref\wadia{G. Mandal, A. M. Sengupta, and S. R. Wadia, Mod.~Phys.~Lett.~A6
(1991) 1685.}
\nref\jevicki{J. Avan and A. Jevicki, Phys.Lett. 226B (1991) 35 and 272B (1991)
17.}
\nref\morepol{D. Minic, J. Polchinski and Z. Yang,  Nucl.Phys.B369 (1992) 324.}
\nref\morewadia{A. Dhar, G. Mandal and S. R. Wadia, Mod.Phys.Lett. A7 (1992)
3129 and A8 (1993) 3557.}
\nref\dmp{R. Dijkgraaf, G. Moore and R. Plesser, Nucl.Phys. B394 (1993)
356-382.}
\nref\witten{E.~Witten, Comm.~Math.~Phys.~121 (1989) 351.}
\nref\elitzur{S.~Elitzur, G.~Moore, A.~Schwimmer and N.~Seiberg,
Nucl.~Phys.~B326 (1989) 108.}
\nref\witetc{S.~Axelrod, S.~Della Pietra, E.~Witten,
J. Diff. Geom. 33 (1991) 787;\Cr S. Axelrod, Ph.D. Thesis, Princeton University
(1991).}
\nref\bernard{D. Bernard, Nucl.~Phys.~B303 (1988) 77.}
\nref\gawedzki{K.~Gawedzki, Nucl.~Phys.~B328 (1989) 733.}
\nref\geras{A. Gerasimov, hep-th/9305090}
\nref\blauthom{M. Blau and G. Thompson, Nucl.~Phys.~B408 (1993) 345;\Cr
also hep-th/9310144 and hep-th/9402097.}
\nref\witgr{E.~Witten, `The Verlinde Algebra and the Cohomology of the
Grassmannian,' IASSNS-HEP-93-41, hep-th/9312104.}
\nref\gt{D. J. Gross and W. Taylor, Nucl. Phys. B400 (1993) 181 and
 Nucl. Phys. B403 (1993) 395.}
\nref\cmr{S. Cordes, G. Moore and S. Ramgoolam,
``Large N 2D Yang-Mills Theory and Topological String Theory,''
YCTP-P23-93, RU-94-20, hep-th/9402107.}
\newsec{Introduction}
Most of the solvable large $N$ models which have been related to string
theories, in particular the $c\le 1$ matrix models and two-dimensional
Yang-Mills theory, boil down in a formal sense to free fermions.
The essential observations
go back to Weyl and Harish-Chandra, but were first systematically exploited in
this context by Brezin et. al. \BIPZ~
One can evaluate the inner product on singlet wave functions using the Weyl
integral formula; this produces a measure factor which can be absorbed into the
wave functions, making them totally antisymmetric; furthermore this
redefinition also turns
natural Hamiltonians into free Hamiltonians.
This solves the theory for any $N$ but is particularly useful in the large $N$
limit -- the ground state is that of a one-dimensional Fermi liquid and
completely described by its Fermi surface, and observables are naturally
described in a second quantized formalism which can be thought of either as a
quasi-relativistic Fermi system or an interacting bosonic system.
Among the many papers which develop this formalism we mention
\refs{\bosecone,\polchinski,\wadia,\jevicki,\morepol,\morewadia,\dmp}.

A prototypical topological field theory is $D=3$
gauge theory with a pure Chern-Simons action, as solved by Witten \witten~
(and referred to as `CSW theory' in the following).
In this note we describe
a free fermion formulation of the Hilbert space for CSW theory with gauge group
$U(N)$ on space-time $T^2\times I$.
Although CSW theory has no Hamiltonian, we will argue that the other elements
of the picture apply.
The only real difference is that momentum space is periodic, and there are a
finite number $N+k$ of discrete momenta.

\nref\toappear{M. Douglas, to appear.}
The quantization of CSW theory has been much studied
\refs{\witten,\elitzur,\witetc}
and we will quote from these works and work on the closely related
WZW model \refs{\bernard,\gawedzki} and
$G/G$ gauged WZW model \refs{\geras,\blauthom,\witgr} to justify our picture.
%leaving a more formal derivation for \toappear.
Ideally this would be a simple application of the `Weyl integral formula for
path integrals' already implicit in these works and proposed more explicitly in
\blauthom.
Our results can also be considered as a
%simplified derivation (using existing results for CSW theory) and
re-interpretation of the description of the
$SU(N)$ Verlinde algebra due to Gepner.
\ref\gepner{D. Gepner, Comm.~Math.~Phys.~141 (1991) 381.}
Actually we discuss a slightly simpler case,
the fusion ring of $U(N)$ at level $(k,N(k+N))$.
\ref\intr{K. Intriligator, Mod.Phys.Lett.A6 (1991) 3543.}
This appeared in Witten's recent work \witgr~
relating the Verlinde algebra to the
quantum cohomology of a Grassmannian sigma model.

Though the fermionic description is valid at finite $N$ and $k$,
our original motivation for this work was to study
the large $N$ limit of CSW theory, and explore the possibility of
duplicating the program of \gt, where a string interpretation was derived for
two-dimensional Yang-Mills theory on arbitrary genus surfaces.
We will discuss this in the second part of the paper.
The large $N$ limit of CSW theory was first studied by Camperi, Levstein and
Zemba.
\ref\clz{M. Camperi, F. Levstein and G. Zemba, Phys. Lett. B247 (1990) 549.}
Exact results can be found for the partition function on simple three-manifolds
by exploiting the relation with conformal field theory \witten,
and in \clz~ the large $N$ expansion of $Z(S^3)$ was found and compared with
expectations from general large $N$ considerations.
Further results were obtained by Periwal
\ref\peri{V. Periwal, Phys.~Rev.~Lett.~71 (1993) 1295.}, namely the expansion
of free energies on $S^3$ and $T^3$.
He pointed out that one might expect a topological closed string representation
analogous to that of \gt, and showed that these exact results exhibit striking
similarities with other low-dimensional string results, such as an equality
between the $O(N^{2-2g})$ term in the free energy on $S^3$ (hypothetically the
result of a path integral over world-sheets of genus $g$ embedded into $S^3$)
and the Euler characteristic of genus $g$ moduli space.
The full picture still remained murky, and there are as yet unexplained
differences with known large $N$ limits, for example the free energy of $T^3$
has an expansion in odd powers of $1/N$.

\nref\polym{J. A. Minahan and A. P. Polychronakos,
Phys. Lett. B312 (1993) 155.}
\nref\douglas{M. R. Douglas, Proceedings of the 1993 Carg\`ese workshop on
Strings and Quantum Field Theory, hep-th/9311130.}
The fermionic formulation makes the large $N$ limit simple to describe.
Holding the parameter $x=N/(k+N)$ fixed,
CSW states will become
configurations of a classical fermion liquid with a two-dimensional torus as
phase space.  These
can typically be described by their Fermi surface.
The algebra of Wilson loops contains a $W_\infty$ algebra, and
general Wilson loops (which are contained in a finite region $T^2\times I$ and
so can be regarded as operators on $H(T^2)$) can be rewritten using it,
allowing their correlation functions in the large $N$ limit to be calculated
using existing techniques.
Modular transformations on $T^2$ also act simply and we discuss
$Z(S^3)$ from this point of view.
{}From this one can see that there is no double-scaling limit of CSW theory.

The formal structure is very similar to YM$_2$ and thus one has a reason to
believe the same string reformulation will work here.
We review the bosonization approach of
\refs{\polym,\douglas}
and state a hypothesis which might help guide future work on string
reformulations: namely, that if a field theory has a string reformulation, its
Hilbert space will also have a string reformulation.
In higher dimensional gauge theories, establishing (or refuting) this might be
possible with perturbative techniques.
Our discussion here is a bit general but is intended to suggest new directions
for research.

Although we will duplicate what for YM$_2$ was a fairly direct path to a
string formulation, giving the CSW results a string interpretation produces
some unusual features.  One can think of Wilson loop operators as modes of a
`classical string field' which as one would expect for a topological string are
functions
on $\pi_1(T^2)$.  The symplectic structure for this string field following from
CSW theory is unusual in the string context; loops around the $a$ and $b$
cycles will be conjugate.  To get a string theory which reproduces $1/N$
corrections, we must work with the quantum theory, and states of this theory
will be given by specifying the occupation numbers of strings winding about
(say) the $a$ cycle only.
This reformulation is not valid for quantities which involve sums
over the entire Hilbert space, the simplest example of which is $Z(T^3)$.
All this means it is not at all obvious whether the
program of \gt\ can be duplicated here.

It is interesting that CSW theory can be derived from a topological open string
theory. \ref\witopen{E. Witten, hep-th/9207094.}
There is an analogy with QCD, which can also be derived from an open string
theory (giving a gauge-fixed perturbative formulation), and which
hypothetically can be reformulated as a closed string theory describing only
gauge-invariant observables.

We will be taking the large $N$ limit of finite $N$ results, but it would be
very interesting to reformulate the theory directly in terms of invariants in
the large $N$ limit (as is done in collective field theory
\ref\js{A. Jevicki and B. Sakita, Nucl.Phys. B165 (1980) 511.}).
To some extent this can be done, but so far we were not able to derive the
shift $k\rightarrow k+N$ in this approach.

There are loop equations for CSW theory, studied in
\ref\awada{M. Awada, Comm.~Math.~Phys.~129 (1990) 329.}.  Our opinion is that
the existence of loop equations is a necessary but not sufficient condition for
a string theory reformulation of a field theory to exist.  The $D=3$ Ising
model, which would have to be a non-topological (in space-time) string, but
simply has too few degrees of freedom, is an illustration.  One needs as well
an exact reformulation of the Hilbert space in string terms, as exists for the
matrix models and YM$_2$ (to all orders in $1/N$).
\newsec{$U(N)$ Chern-Simons theory at genus one}
A CSW theory is specified by gauge group $G$ and integer $k$, and has action
\eqn\csact{S =
{k\over 4\pi}\int_M \Tr(B\wedge dB + {2\over 3}B\wedge B\wedge B)}
with $B$ a gauge connection.
This needs no metric on the three-manifold $M$ for its definition.
In \witten\ it was shown that no metric is needed to regulate the theory,
and that the natural observables, the partition function and expectation values
of closed Wilson loops $L_i$ on $M$, depend only on the topology of
$M - \sum L_i$, the representations $R_i$ taken for the Wilson loops,
and a few further discrete choices of `framing'.
We will generally deal with the framing by making simple canonical choices
possible for the manifold $T^2\times I$ and our observables.

We will take as gauge group $U(N) \cong SU(N)\times U(1)/\BZ_N$.
The gauge couplings for the two factors can of course be chosen independently
and we will make use of this later.
%\eqn\uone{S' = {1\over 4\pi}\int_M \Tr B\wedge\Tr dB}
Eventually, to weigh the diagrammatic expansion by $N^\chi$, we will take
$k=1/g^2$ proportional to $N$.
Consider a region of space-time isomorphic to $\Sigma\times I$, with $\Sigma$ a
two-dimensional Riemann surface and $I$ a one-dimensional interval.
Taking $B_0=0$ gauge, and letting $A$ be the gauge field on $\Sigma$, the
action becomes
\eqn\cscan{S =
-{k\over 4\pi}\int_{\Sigma\times I}~
 \Tr A \ppt A ~dt +
 {k\over 2\pi}\int_{\Sigma\times I}~ \Tr A_0 F}
with $F = dA + A^2$.
The integral over $A_0$ sets $F$ to zero, so the classical phase space is the
space of flat connections on $\Sigma$ modulo gauge transformations.
These are completely determined by their holonomy around the non-contractible
based loops of $\Sigma$, in other words by a group homomorphism
$\pi_1(\Sigma)\rightarrow G$.
The remaining gauge transformations then act on this by the adjoint action.

Following \elitzur, we first describe the naive quantization of the reduced
classical phase space.
As is well known, this procedure is not correct in detail: it misses quantum
effects responsible for (among other things) the famous shift $k\rightarrow
k+N$ in many formulas.
The large $N$ limit is taken with $k\sim N$ so this is important; we merely
summarize the correct treatment here, generally following \elitzur.
An interesting lesson for large $N$ can be drawn from this:
although we expect the limit to be a classical theory,
very basic elements of this theory,
such as the equal-time algebra of observables,
can be different from the original $\hbar\rightarrow 0$ classical theory.

We will refer to the quantum Hilbert space on the surface $\Sigma$ with
insertions of time-like Wilson lines in the representations $R_1, R_2, \ldots$
as $H(\Sigma;R_1, R_2, \ldots)$.
Consider $\Sigma=T^2$ with coordinates $\sigma^a\cong\sigma^a+1$
and $\sigma^b\cong\sigma^b+1$.
Here, the general solution of $F = 0$ is
\eqn\tflat{\tilde A = - \tilde d U U^{-1} + U \theta(t) U^{-1}}
where $U$ is single-valued and $\theta(t)$ is a Lie-algebra-valued one-form
which depends only on $t$.
The homotopy group $\pi_1(T^2)\cong \BZ\oplus\BZ$ is commutative,
so we can choose $U$ to diagonalize both components of $\theta(t)$.
`Large' gauge transformations
(not connected to the identity)
then take $\vec\theta_i$ to
$\vec\theta_i+2\pi \vec n_i$; they also include permutations.
Thus the phase space is $T\times T/W$, where $T\cong (S^1)^N$ is the maximal
torus of $U(N)$, and $W\cong S_N$ is the Weyl group whose elements
simultaneously permute the two components of $\vec\theta_i$.
The change of variables to $\vec\theta_i$ produces a linear measure, and
substituing \tflat~ into the action finally produces
\eqn\effact{S = {k\over 2\pi}\int \sum_{i=1}^N \theta^a_i \dot\theta^b_i}
and the commutation relations
\eqn\torcom{[\theta^a_i,\theta^b_i] = -i{2\pi\over k}\delta_{ij}.}
This free system is easily quantized by choosing (say)
$\theta^a_i$ to play the role of positions and taking
$\theta^b_i=2\pi i{1\over k}\partial/\partial\theta^a_i$.
Since position is compact, the momenta are quantized, and wave functions are
superpositions with $\vec\lambda\in\BZ^N$ of
\eqn\defpsi{\psi_{\vec\lambda} = \exp i\vec\lambda\cdot\vec \theta^a.}

The compactness of the momenta now implies that the number of quantum states is
finite, and it is useful to phrase this as follows.
Shifts $\theta^a\rightarrow\theta^a+\alpha$ and
$\theta^b\rightarrow\theta^b+\beta$ (for each $i$) generate
a Heisenberg-Weyl group:
\eqn\hwgroup{\eqalign{
S^a_\alpha[\psi](\theta) &= \psi(\theta+\alpha)\cr
S^b_\beta[\psi](\theta) &= e^{i k\beta \theta/2\pi } \psi(\theta)\cr
S^a_\alpha S^b_\beta \psi &=
e^{ik \alpha \beta/2\pi } S^b_\beta S^a_\alpha \psi.}}
The commutator implies that we can only impose simultaneous periodicity with
$\alpha \beta=4\pi^2 n/k$, $n\in\BZ$, so $S^a_{2\pi}=1$ is compatible with
$S^b_{2\pi n/k}=1$.
The large gauge transformation is $S^b_{2\pi}$ which is $k$ fundamental units,
and in terms of the momenta $\lambda$ this is $\lambda\cong\lambda+k$.
Thus we can implement the constraint by superposing wave functions
$\psi_{\vec\lambda+k\vec n}$, and our state space has a basis labelled by
$\lambda\in \BZ^N/W\times k\BZ^N$.

This analysis is correct qualitatively, but not quantitatively.
In a correct treatment, one must integrate out the non-constant modes of the
gauge field, which will produce an effective theory very similar to the above
but with a finite renormalization of the parameter $k$.
This can already be seen in a careful perturbative treatment
(since the theory is finite).
\nref\aglr{L. Alvarez-Gaum\'e, J.~M.~F. Labastida and A.~V.~Ramallo,
Nucl.~Phys.~B334 (1990) 103.}
\refs{\witten,\aglr}
The existing derivations \refs{\elitzur,\witetc,\blauthom} of this
are rather intricate, and we only try to give the essential idea here.
(The result, for example \elitzur, equations (4.12) and (4.13), is much simpler
than the derivation.)
To integrate out the non-constant modes in a well-defined way,
one must work with holomorphic quantization.
Define a complex coordinate $z=\sigma^a+\tau\sigma^b$.
Wave functions will be holomorphic functions of the complex gauge field
$A_z=(A_2-\bar\tau A_1)/(\tau-\bar\tau)$, which will again be reduced to its
zero modes $a_i=A_{\bar z}/2\pi(\tau-\bar\tau)$,
for which the inner product becomes
\eqn\holoip{\vev{\chi|\psi} =
\int \prod_i d^2a_i~ e^{-k(\im a)^2\over\im\tau} \chi(a)^* \psi(a).}
This representation is precisely equivalent to the usual one
(by the Stone-von Neumann theorem) -- the relation is
\eqn\posbas{\psi(a)=\int d\theta~
e^{-{ik\over 2\tau}(\theta-2\pi a)^2} \psi(\theta)}
and the Heisenberg-Weyl group \hwgroup\ is transformed to
\eqn\hwholo{\eqalign{
S^a_\alpha[\psi](a) &= \psi(a+\alpha/2\pi)\cr
S^b_\beta[\psi](a) &= e^{-i k \tau\beta^2/8\pi^2 +i k\beta a }
\psi(a+{\tau\beta\over 2\pi}).}}
We will use this the same way we did earlier:
impose invariance under the large gauge transformations
$S^a_{2\pi}$ and $S^b_{2\pi}=S^b_{2\pi n/k}$ with $n=k$.
The second condition is usually stated in another way:
we can think of invariance under $S^b_{2\pi }$ as identifying $\psi(a)\cong
\psi(a+\tau)$ up to a transition function $e^{2\pi ika}$, non-trivial because
it depends on $a$.
Since $a$ parameterizes flat connections, we say that the wave function is a
section of a holomorphic vector bundle of degree $k$ over the space of flat
connections.  \witten

To reduce a general wave function of $A_z(z)$ to the zero modes,
we follow \elitzur\ and write an inner product of two wave functions
\eqn\csip{
\vev{\chi|\psi} = \int D\bar A DA~e^{-{ik\over 2\pi}\int \Tr A\bar A}
(\chi[A])^*~\psi[A].}
This will be computed by changing variables from $A$
to $\theta$ and the `complexified
gauge group' $g(z,\bar z):\Sigma\rightarrow G^\BC$.
It can be seen that
\eqn\ginv{\psi[A_z]|_{A_z=g^{-1}(\partial_z+\theta_z)g}~
 =~ e^{-ikS_{GWZW}(g,\theta_z,0)} \psi[\theta_z]}
(where $S_{GWZW}(g,A,\bar A)$ is the gauged WZW action)
is a gauge-invariant wave function of $A_z$.
%(In \elitzur~ this is done for $\bar A$ as well.)
Computing the Jacobian for this change of variables and integrating out $g$ and
$\bar g$ then produces
\eqn\finip{\vev{\chi|\psi} = \int d^2a~
e^{-{(k+N)(\im a)^2\over \im \tau}} |\Pi(\tau,a)|^2 (\chi[a])^* \psi[a]}
where \footnote*{Here $q=\exp {2\pi i\tau}$, $z_i=\exp 2\pi i a_i$,
$({\rm Ad}[z_i])_{jk}\equiv z_j/z_k$ is the adjoint representation of the group
and $\tilde\Delta(z)=\det (1-{\rm Ad}[z])^{1/2}
=\prod_{i<j}(z_i-z_j)/\prod z_i^{(N-1)/2}$.}
\eqn\defpi{
\Pi(\tau,\theta) = \tilde\Delta(z) \prod_{n\ge 1}
\det (1-q^n {\rm Ad}[z])}
is the denominator of the Weyl-Kac character formula.
\ref\kac{V. G. Kac, {\it Infinite-Dimensional Lie Algebras}, Cambridge Univ.
Press, 1985.}
The calculation of $\Pi$ is relatively straightforward, while the prefactor
requires some care with the zero modes.
$\Pi$ is antisymmetric under the Weyl group and behaves simply under the large
gauge transformations $a_i\rightarrow a_i+1$ (under which $\Pi$ is invariant)
and
$a_i\rightarrow a_i+\tau$, under which we have
\eqn\mshift{
\Pi[z_1,\ldots,z_i\rightarrow qz_i,\ldots] = (-1)^{N-1} {\prod_j z_j\over
z_i^N} \Pi[\vec z].}

We now want to treat this formula in the same spirit as the Weyl integral
formula, and redefine the wave functions $\Pi\psi\rightarrow\psi$ to make the
inner product trivial.
They will then be completely antisymmetric under the Weyl group, and they will
be holomorphic.  Imposing invariance under the large gauge transformations will
make them sections of a holomorphic vector bundle, but they will no longer have
degree $k$, because of the transformation law
\mshift.  From this and the comments below \hwholo\ we see that $\Pi$
transforms as a section of a bundle of degree $N$, and the redefined wave
functions will be sections of a bundle of degree $k+N$.
($\Pi$ as well as the shift in the prefactor only see the $SU(N)$ subgroup, we
will duplicate this
for the $U(1)$ factor by hand below.)
These constraints have a finite number of solutions, the numerators in the
Weyl-Kac formula for integrable representations.  They can be simply written in
terms of theta functions.

The result is that the redefined wavefunctions are those for a free system of
$N$ fermions much like our original treatment but with $\hbar=2\pi/(k+N)$.
Furthermore the large gauge transformations are also $S^a_{2\pi}$ and
$S^b_{2\pi(k+N)/(k+N)}$ as in \hwholo\ with $k\rightarrow k+N$.
It is now straightforward to go back to the `position basis', as in \elitzur.
We will argue that certain observables act simply in this basis,
but in fact we will describe the action of observables in terms which can be
immediately translated into the action of the Heisenberg-Weyl group,
so \hwholo\ defines their action on the holomorphic wavefunctions.

The commutation relations become
\eqn\truecom{[\theta^a_i,\theta^b_i] =
-i{2\pi\over k+N}(\delta_{ij}-{1\over N}\delta_i\delta_j)
-i{2\pi\over k'}\delta_i\delta_j}
where we explicitly decomposed into the $SU(N)$ factor with coupling $k$ and
the $U(1)$ factor with coupling $k'$.
Now we will use our freedom to choose the $U(1)$ coupling $k'=N(k+N)$ to
duplicate the shift there.
The wave functions periodic in each variable in position space are
$\psi_{\vec\lambda}$ with $\vec\lambda\in(\BZ+[n_F])^N$ on the $U(N)$ weight
lattice, where $n_F=(N-1)/2$ and $[n_F]$ is its fractional part.
In using this weight lattice we have already quotiented by $\BZ_N$ acting on
the holonomy $U^a$.
The $1/2$ for $N$ even comes from absorbing
$\tilde\Delta$ into the wave function,
as in $U(N)$ group quantum mechanics \douglas.
Here it amounts to a convenient choice of phase convention for wave functions,
simplifying some formulas.
The periodicity of momentum space
is implemented by superposing wave functions
$\psi_{\vec\lambda+(k+N)\vec n}$
with $\vec n\in \BZ^N$.
In doing this we have implemented the quotient by $\BZ_N$ acting on the
holonomy $U^b$, which will also mix $SU(N)$ and $U(1)$ as we will see.
Explicitly summing the superposed wave functions will produce delta-functions
constraining the $\theta^a$ to satisfy
$\exp i(k+N)\theta^a = e^{i\omega}$.
The choice of the phase $e^{i\omega}$ is a convention at this point;
however the expressions for the observables will certainly depend on it.
The symmetry between $\theta^a$ and $\theta^b$
suggests the choice $e^{i\omega}=(-1)^{N-1}$
which has solutions $\theta^a\in 2\pi(\BZ+[n_F])/(N+k)$.
Finally we mod out by the Weyl group $W\cong S_N$ by completely
anti-symmetrizing the wave functions.
It is well known that the CSW Hilbert space on $\Sigma=T^2$ has a basis
labelled by
\eqn\deflam{\lambda\in \tilde\Lambda^{(k)}\equiv
\left({\Lambda^w\over W\times (k+N)\Lambda^R}\right)^{\#}}
where the notation $(\ldots)^{\#}$ indicates that fixed points under the group
action are removed, and of course this is exactly accomplished by
antisymmetrizing the wave functions.

The states correspond to irreducible representations of
$U(N)$ essentially as for group quantum mechanics, \douglas~
with $\lambda=\alpha+\rho$ the shifted highest weight.
The periodicity in momentum allows a finite subset of them, the integrable
representations of the level $k$ affine algebra.
We can also translate into the language of Young tableaux.
The trivial representation is $\lambda=\rho$, a state $\ket{0}$ with levels
$-n_F\le i\le n_F$ filled.
A representation whose tableau has $n_i$ boxes in the $i$'th
row corresponds to moving the $i$'th fermion (counting from the
top) up $n_i$ levels.  Since there are $N$ fermions and $N+k$ levels, the
constraint in momentum space should correspond to keeping Young tableaux
with at most $k$ boxes in a row.
This is the correct constraint on integrable representations of $SU(N)_k$,
of which there are $N-1+k\choose k$.
Our Hilbert space has dimension $N+k\choose k$,
which differs by a factor $(N+k)/N$.  This is $N(N+k)$, the number of states of
the $U(1)$ Chern-Simons Hilbert space, divided by the volume of our discrete
symmetry $1/N^2$.  One $1/N$ came from the correlation between $SU(N)$ and
$U(1)$ weights produced by using the weight lattice $\BZ^N$.
The other $1/N$ comes because we do not distinguish the $U(1)$ fermion from the
$SU(N)$ degrees of freedom.  Thus our free fermion Hilbert space will identify
(for example) the irreps (labelled by $0\le h<N$) with $U(1)$ charge $-hk$ and
rectangular Young tableau of width $k$ and height $h$.

The basic observables are the Wilson loops $W_C[R]=\Tr_R\exp i\int_C dx^i A_i$
integrated over a closed contour $C$ and with the trace taken in a
representation $R$.
Eventually we will want to specify a representative of each homotopy class
defined without regard to time ordering, for example
$W_{n_a,n_b}\sim\Tr U_a^{n_a}~U_b^{n_b}$ in terms of holonomies in the
fundamental representation
$U_i=\exp i\int dx^i A_i$.
To embed such a loop in $T^2\times I$ without self-intersection we must choose
an embedding in time as well and there are many possibilities, even for the
simplest cases $W_{n_a,0}$ and $W_{0,n_b}$.  Thus it is simpler to start with
$W_a[R]=W_{1,0}[R]$
and $W_b[R]$ which are unambiguous.
In a proper reduction to the global variables $\theta_i^a$,
it is necessary to regulate the Wilson loop by a form of `point splitting'
\witten, in which the self-energy of a loop is defined by choosing a vector
field $v(s)$ normal to the loop, and evaluating self-energies between $x(s)$
parameterizing $W$ and $x(s)+\epsilon v(s)$.  The choice of $v(s)$ is referred
to as a framing, and all results depend on the winding number of $v(s)$.  Since
we will only consider $\Sigma=T^2$, we will simply work with the convention
that these vector fields are always chosen to point along the time-like
direction.  The large $N$ limit of the framing dependence, from
\witten, will be non-trivial: a $2\pi$ twist acts on the fundamental
representation as $e^{2\pi iN/(k+N)}$.

Consider the Wilson loops
\eqn\wilR{W_a[R] = {\rm Tr}_R U_a = \chi_R(U_a).}
They correspond directly to the states $\ket{R}$ as follows: \witten~
do the path integral over a solid torus $S^1\times D^2$, with $W_a[R]$ inserted
at the center of the two-disk $D^2$ (so $a$ labels the non-contractable cycle);
then the boundary wave functional with $\theta^a$ as positions is $\ket{R}$.
The inner product between two states $\vev{R|R'}=\delta_{R,R'}$ and can be
re-interpreted as the trace over the Hilbert space $H(S^2;R,R')$.

We assume, following \elitzur,
that these loops have a direct translation to the position basis
\eqn\wilRp{W_a[R] = \chi_R(e^{i\theta^a}) =
{1\over|R|!}\sum_{\sigma\in S_{|R|}} \chi_R(\sigma)
\prod_i P_{\sigma_i}[e^{i\theta^a}]}
with the last equality being the Frobenius relation.
In the fermionic formalism the power sums $P_n$ are bilinears
\eqn\powersum{
P_{n}[e^{i\theta^a}]=\sum_i e^{in \theta^a_i}
= \sum_{m\in L} B^+_{-n-m} B_m}
in a (standard) second quantized notation where $B^+_{-n}$ creates
the mode $e^{in\theta^a}$,
$\{B^+_{m}, B_n\}=\delta_{n+m \bmod N+k,0}$ and
$L\equiv \BZ/\BZ_{N+k}+[n_F]$.

Now we can compute the fusion (Verlinde algebra) coefficients
\eqn\fusion{N_{RST} = \vev{0|W_{R_a}W_{S_a}W_{T_a}|0}.}
Recall \douglas\ that the state corresponding to a representation labelled by a
Young tableau with $n_i$ boxes in the $i$'th row can be written
\eqn\fermstate{\prod_{i=1}^N B^+_{-(n_F+1-i+n_i)} \ket{}}
where $\ket{}$ is the state with fermion number zero; then using the Frobenius
relation for the third representation produces a fermionic expression which
makes the truncations due to finite $k$ quite manifest.
Rank-level duality $(N,k)\rightarrow (k,N)$
\ref\rank{E.J. Mlawer, S.G. Naculich, H.A. Riggs, and H.J. Schnitzer, \Cr
Nucl.~Phys.~B352 (1991) 863.}
is also manifest -- in terms of a basis labelled by fermion occupation numbers
$\ket{\{n_i\}}$, the state $\ket{\{1-n_i\}}$ is a state of the dual theory, and
all observables in the two theories are simply related by the exchange
$B^+_n\leftrightarrow B_{n}$.
This transposes the Young tableaux but also prescribes signs and the
treatment of the $U(1)$ factor.

This description of the fusion ring
could also have been derived quite directly from the results of
\refs{\gepner,\witgr}.
There it is shown that the fusion ring of $U(N)_{(k,N(N+k))}$ can be realized
as the ring of symmetric polynomials in $N$ variables $\lambda_i$ satisfying
\eqn\witcon{\lambda_i^{N+k} = (-1)^{N-1}.}
There is a functional on this ring $J(f)$ which can be used to
define an inner product $(f|g)=J(f^* g)$.
It is a sum over the set of $\lambda_i$ satisfying \witcon.
It is given by (3.45) in \witgr: after a shift of the overall $U(1)$ charge to
zero
%(as in (4.63))
it is
\eqn\defJ{J(f) = {1\over N!}
\sum_{\lambda_i} \prod_{i<j}|\lambda_i-\lambda_j|^2 f(\lambda).}
The Vandermonde forces the $\lambda_i$ in the sum to be distinct, and
clearly if we absorb the factor $\prod_{i<j}(\lambda_i-\lambda_j)$ into our
`wave functions' $f$ and $g$ we will reproduce the $N$ fermion Hilbert space we
found with $\lambda_i=e^{2\pi i\theta^a_i}$.  We also confirm the compatibility
of our choice $\theta^a\in 2\pi L/(N+k)$ with \wilRp.

We believe this presentation of the fusion ring should be derivable entirely
from the CSW path integral, and the presentation of the fusion ring in
\blauthom\
is very close to this.
All we have done here is to identify it with the natural action of Wilson loops
on $H(T^2)$, and apply second quantization.

In the fermionic formalism it is natural to consider all
the one-particle operators
\eqn\willoop{\eqalign{
W_{n_a,n_b}&=\sum_i e^{i(n_a \theta^a_i + n_b \theta^b_i)}\cr
&= e^{-\pi i n_a n_b /(N+k)} \sum_{m\in L}
 e^{-2\pi i n_b m/(N+k)} B^+_{-n_a-m} B_m}.}
These have commutation relations
\eqn\fincom{[W_{m,n},W_{m',n'}] =
2i \sin \left({\pi(m n'-m' n)\over k+N}\right)~
W_{m+m',n+n'}.}
At leading order in $1/N$ we can identify these as $W_{n_a,n_b}=\Tr
U_a^{n_a}~U_b^{n_b}+O(1/N)$.  For the conclusions we draw regarding a string
interpretation, this identification will suffice, and the following two
paragraphs are not essential.

The exact relation at subleading orders or finite $N$ seems rather subtle.
We can compare with the skein relation of \witten\ which with our framing
conventions is
\eqn\skein{[L_{1,0},L_{0,1}] = 2i \sin {\pi\over k+N} L_{1,1}.}
(The notation here is to draw the same picture as for the corresponding loops
$W_{m,n}$ but then interpret the relation as a local commutation relation at
the crossing.)
The basic commutator $[W_{1,0},W_{0,1}]$ is simple, while
consideration of the general case leads to the conclusion that the relation of
these operators to Wilson loops defined with simple paths such as $\Tr
U_a^{n_a}U_b^{n_b}$ with time ordering corresponding to operator ordering is
not completely trivial.  The realization that there is no canonical time
ordering for these operators should make this less surprising and
in writing ``$W_{n_a,n_b}=\Tr U_a^{n_a}~U_b^{n_b}$''
we have oversimplified the real situation.  If we compute the commutator of two
such Wilson loops, the procedure of re-ordering the result into this form
will produce non-linear terms.
Thus, although all observables which act on $H(T^2)$ and produce a state in
$H(T^2)$ can be written in the fermionic formalism, it is not immediately
obvious which ones are bilinears and which are not.

The $W_{n_a,n_b}$ can be written as linear sums of connected fundamental
representation Wilson loops.  (If we allow ourselves to use higher
representations, they are much easier to define using \wilRp).
The precise relation
is most easily made by starting with $W_{1,0}$, which is trivial, and
its images under modular transformations, which take the $(a,b)$ cycles
to $(wa+xb,ya+zb)$.  (We will say more about these below in the large $N$
limit.)
Homotopy classes $(p,q)$ fall into orbits labelled by `total winding number'
$|\gcd(p,q)|$ and a representative in each orbit can be obtained by commuting
two loops of total winding number $1$, for example $[W_{1,n},W_{-1,0}]$.
Since modular transformations act independently on each of the $N$ fermions,
acting on $W_{1,0}$ we verify that $W_{1,n}$ as defined by \willoop\ is
also a single connected Wilson loop.
Now let \fincom\ define the general $W_{p,q}$.
To get its expression as a sum of connected, time ordered Wilson loops,
we compute the same commutator using the skein relation.  If we refrain from
further reordering of the result, the result is a sum of connected Wilson
loops.  It would be interesting to get a precise expression for the relation.

\smallskip
The algebra \fincom\ is isomorphic to the Lie algebra $SU(N+k)$.
\ref\fairlie{D. Fairlie, P. Fletcher and C. Zachos, Phys.~Lett.~B218 (1989)
203.}
Our fermionic Hilbert space $H(T^2)$ decomposes into the sum of totally
antisymmetric
$SU(N+k)$ irreps (labelled by the fermion number $N$), and the modular group
action factors through $SU(N+k)$.
Amusingly enough, the orbit  $\exp t_{mn} W_{mn}\ket{0}$ is a Grassmannian
$G(N,N+k)$ isomorphic to that considered in \witgr.
The complete CSW theory does not have $SU(N+k)$ symmetry, however, and the full
significance of these observations is not clear to us.
\newsec{The large $N$ limit}
In section 2, we found that the Hilbert space $H(T^2)$ is a truncation of that
for group
quantum mechanics: both `position' and `momentum' variables are compact,
since both conjugate variables came from components of a gauge connection.
The observables also have a simple relation to those of group quantum
mechanics.

The large $N$ limit of this phase space is well-known
and can be treated exactly as was the matrix model: \refs{\polchinski,\wadia}
since $2\pi/(k+N)\sim \hbar$, the system becomes classical.
Let us define $q_i=\theta^a_i/2\pi$ and
$p_i=\theta^b_i/2\pi=-i{1\over k+N}\partial/\partial\theta^a_i$,
which become classical in the large $N$ limit.
The CSW phase space is then the space of phase space fermion densities
$\rho(q,p)$,
where at each point $\rho$ can take only the values
$0$ or $1$.
It will satisfy
\eqn\normal{\int dq dp~\rho(q,p) = {N\over k+N} \equiv x.}
As in \refs{\clz,\peri}, particular large $N$ limits should be characterized
by the parameter $x$.
Typical applications of matrix quantum mechanics involved states in which the
Fermi surface (the boundary between $\rho(q,p)=0$ and $1$) was a simple smooth
curve.

\nref\dk{M.~R.~Douglas and V.~Kazakov, Phys.~Lett.~B319 (1993) 219.}
As in \dk~ these densities $\rho$ can be thought of as characterizing
particular `master' representations in the large $N$ limit.
There the basic variable was $u(p) = \int dq~ \rho(q,p)$.
The compactness of $q$ showed up in the bound $0\le u(p)\le 1$.
Using this correspondance, we see for example that the trivial representation
$\ket{0}$ corresponds to the phase space density
\eqn\trivial{\rho_0(q,p) = \cases{1\qquad |p|\le {x\over 2}\cr
0\qquad {\rm otherwise.}}}

The Wilson loops are the basic variables of a string representation, and we
expect that in the large $N$ limit their expectation values would form a good
set of coordinates on phase space.  Evidently they will become
\eqn\wilqp{\eqalign{\vev{W_{n_a,n_b}}&= \sum_i e^{2\pi i(n_a q_i + n_b p_i)}\cr
&\rightarrow N\int dqdp~\rho(q,p) e^{2\pi i(n_a q + n_b p)}\cr
&\equiv (N+k) \vev{w_{n_a,n_b}},}}
the Fourier coefficients of the phase space density $\rho$.
A $\rho$ whose boundary is piecewise continuous
will be determined uniquely by its Fourier coefficients.

The simplest case is to take $\vev{0|\prod_{i=1}^n w_i|0}$ which will
correspond to
loops embedded in $S^2\times S^1$.
On general grounds, we expect to be able to compute the leading
$O(N^{2-2n})$
connected part of this in a classical `hydrodynamic' formalism.
Since each fermion occupies a fixed volume in phase space,
underlying this formalism is the group action of $\SDiff$.
\ref\arnold{V. Arnold, {\it Mathematical Methods of Classical Mechanics},
app. 2.K, Springer 1978.}
Its Lie algebra is the large $N$ limit of \fincom~ (the
`$w_\infty$ algebra' discussed by many authors):
\eqn\wcom{\{w_{m,n},w_{m',n'}\} =
{2\pi \over (N+k)^2}(m n'-m' n)~
w_{m+m',n+n'}.}
In $(p,q)$ coordinates, this is just the Poisson bracket derived from
$\omega = 2\pi dq\wedge dp$.
This determines not only the algebra of observables but also their action on a
state:
\eqn\evolve{w\ket{\rho} = \ket{\rho+{1\over (N+k)^2}\{w,\rho\}}.}
Formally, one can show that the state is in a coadjoint orbit representation.
\nref\yaffe{L. Yaffe, Rev.Mod.Phys. 54 (1982) 407.}
\nref\perelomov{A. Perelomov, {\it Generalized Coherent States and their
Applications},\Cr Springer, 1986.}
\refs{\yaffe,\perelomov,\morewadia}

We postpone an full treatment of this to \toappear\ though the formalism can be
taken over with little change from
\refs{\polchinski,\wadia,\jevicki,\morepol,\morewadia}.
A number of statements follow quite directly.
Loops about the $b$ cycle can be contracted on $D^2$ and
the state $\ket{0}$ is preserved by their action.
Loops about the $a$ cycle will act non-trivially, and
to compute correlation functions, it is useful to have an explicit action $S$
producing this symplectic structure, as in \morewadia.  The problem is then to
find a solution $\rho_{min}(p,q;t)$ with boundary conditions
$\rho|_{t\rightarrow\pm\infty}=\rho_0$ which minimizes
$S + \sum_i w_i(t_i)$.  $S[\rho_{min}]$ is then the generating function
for connected correlation functions.
{}From the discussion of section 2 we saw that much of the topological
information about the embeddings of Wilson loops is lost in the large $N$ limit
so this seems to be of
more interest as the following statement: namely,
if we consider expectation values of Wilson loops in representations with
$O(N)$ highest weight, this construction defines a non-trivial large $N$ limit
of the Verlinde algebra.
For present purposes we will consider the representations with $O(N^0)$
highest weight, and their correlation functions are more simply treated in the
quantum formalism of section 5.

Another operation on $H(T^2)$ is to redefine the $a$ and $b$ cycles, in other
words a modular transformation.  A transformation can be described by defining
the new cycles as linear combinations of the old which have intersection number
one, in other words by an element of $SL(2,\BZ)$.
Its action on a flat connection is simply determined from its natural action on
$\pi_1(T^2)$ and this statement is true in the correctly quantized theory as
well (up to overall phases).
Since we have such a direct correspondance $q=\theta^a/2\pi$ and
$p=\theta^b/2\pi$, the action of the modular group on a state
$\rho(q,p)$ is very simple:  an $SL(2,\BZ)$ element
$E=\pmatrix{a&b\cr c&d\cr}$ acts as
\eqn\classmod{E[\rho] (q,p) = \rho( aq+b p, c q + d p ).}
It produces modular transformations of the phase torus!
These are just elements of $\SDiff$ which are not continuously connected
to the identity.
We illustrate with a picture:
\fig{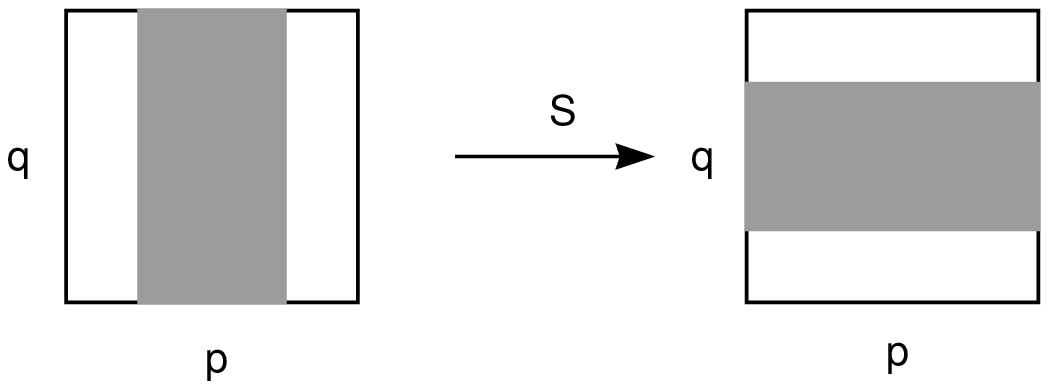}{The modular transform of the trivial representation.}

Now as in \witten\ we can use the modular transformations to compute partition
functions on topologically non-trivial 3-manifolds.
The simplest example is to take two solid tori and glue them together with a
modular transformation, identifying the $(a,b)$ cycles on the first with a
different pair of cycles on the second.
Taking $E=S$, the modular transformation rotating the $a$ into the $b$ cycle,
produces the manifold $S^3$:
\eqn\sthree{Z(S^3) = \vev{0|S|0} = S_{0,0}.}

To evaluate this we need a formula for the overlap of the wave functions
corresponding to two general phase space distributions $\rho_1$ and $\rho_2$.
Since the large $N$ limit is classical this must go to zero for
$\rho_1\ne\rho_2$ as $N\rightarrow\infty$, but what we are interested in is the
free energy:
\eqn\largefree{\vev{\rho_1|\rho_2} = e^{-N^2 F(\rho_1,\rho_2)}.}

Let us consider matrix elements of the modular transformation $S$ between
states each corresponding to integrable representations (so with
$\rho(p,\theta)$ independent of $\theta$).
The finite $N$ result is
\eqn\finmod{S_{R,R'} = C_{N,k} \det_{i,j} s(n_i,n'_j)}
\eqn\modtran{s(m,n) = \sum_{0\le p<N+k\atop 0\le q<N+k} e^{-2\pi ipq/(N+k)}
e^{2\pi imq/(N+k)} e^{2\pi inp/(N+k)}.}
(We will drop the normalization $C_{N,k}$.)
Doing the sum produces
\eqn\finmodtwo{S_{R,R'} = \det_{i,j} \omega^{n_i n'_j}}
with $\omega = \exp 2\pi i/(k+N)$.

The simplest case is where $R'$ is the trivial representation,
because then the $n'_j$ are successive integers, and the determinant reduces to
a Vandermonde:
\eqn\finmodtriv{S_{R,0} = \prod_{i<j} (\omega^{n_i} - \omega^{n_j}).}
For $R$ trivial as well the large $N$ expansion is given in \refs{\clz,\peri}.
We will not reproduce this but simply apply the Euler-MacLaurin formula
\eqn\sthreeexp{\eqalign{\log Z(S^3) &\equiv \sum_{g\ge 0} N^{2-2g} F_g(x)
= \sum_{g\ge 0} (x(N+k))^{2-2g} F_g(x)\cr
&= \sum_{m=1}^{N-1} (N-1-m) \log \sin {\pi m\over N+k} + {\rm trivial}\cr
&= (N+k)^2 \int_{1/(N+k)}^x dy (x-y)\log \sin {\pi y}\cr
&~~+ \sum_{p\ge 1} {B_p\over (N+k)^{p-1} p!} {\p^{p-1}\over\p y^{p-1}}
((x-y)\log \sin {\pi y})|_{y=1/(N+k)}^{y=x}.}}
This could be further simplified but
the only point we want to make about the result is that
at each order in $1/N$, $F_g(x)$
is analytic except at the endpoints $x=0$ and $1$.
Since CSW theory has a perturbative expansion with terms of both signs,
it was not completely clear a priori whether the free energy would have a
singularity coming from summing an exponentially large number of planar
diagrams, and the answer is no: $x=N/(k+N)\rightarrow 0$ is the semiclassical
limit whose singularities cannot be interpreted this way, while $x\rightarrow
1$
in terms of the rescaled coupling $k=N/g^2$ is $g\rightarrow\infty$ which also
does not correspond to an exponential asymptotic for the number of planar
diagrams.
(Even more clearly, the Wilson loop amplitudes on $S^2\times S^1$ had no
singularities in $x$.)
Thus there is no double-scaled string theory,
but there might be a gauge string interpretation analogous to that of YM$_2$.

The large $N$ limit for general $R$ is clearly
\eqn\fourn{F = -\half\int dp dp'~ \rho_R(p) \rho_R(p') \log 2\sin
{\pi|p-p'|\over k+N}.}

The more basic object is the overlap \largefree\ in terms of which any matrix
element of any element of $SL(2,\BZ)$ is determined.
Although the general formula for this does not seem to be in the literature,
existing large $N$ techniques
\ref\largen{C. Itzykson and J.-B. Zuber, J.Math.Phys. 21 (1980) 411;\Cr
A. Matytsin, Nucl.Phys.B411 (1994) 805.}
should provide the large $N$ expansion of the general $F({\rho_1,\rho_2})$.
\toappear

The conclusion is that the reformulation as a
one-dimensional classical fermion liquid suffices to compute the free energy at
$O(N^2)$ and connected Wilson loop expectations at leading order in $1/N$.
Since the variables $W(n_a,n_b)$ have well-defined classical expectation
values, the string interpretation is clearly that these are the components of a
`classical string field' which is a function on $\pi_1(T^2)$, and the
non-trivial structure of this topological
string field theory is the symplectic structure, the action of $SDiff(T^2)$,
and the inner product \largefree.

In a sense the term `classical string field theory' here is only semantic and
as we discuss below
it is not at all clear that the existence of this `classical string field
theory' implies the existence of a `classical world-sheet string theory'
reproducing observables by world-sheet path integrals over genus zero surfaces.

Clearly more work remains to substantiate even this limited sense in which we
have a string theory, and (for example) it is not completely obvious that
the large $N$ limits of the Hilbert spaces $H(\Sigma)$ for higher genus Riemann
surfaces can be thought of as spaces of functions on $\pi_1(\Sigma)$, or what
the analog of the statement that $\rho(q,p)$ was piecewise continuous for
`typical' states might be.

More importantly, our description used a $2+1$ dimensional splitting in a
crucial way, and it would be much more satisfying to remove this dependence.
It is not yet clear to us whether a covariant version of this formalism exists.
Presumably the choices involved in decomposing the original $3$-manifold would
be reflected in structures analogous to those of \gt.  For example, the solid
torus might contain an `$\Omega$-line' winding about the non-contractible
cycle.
\newsec{YM$_2$ strings from a Hamiltonian point of view}
In YM$_2$ it was quite instructive to consider subleading orders in $1/N$,
and in fact a string interpretation could be found for every term in the $1/N$
expansion of the partition function on a Riemann surface, \gt\ with the
contribution of a world-sheet of genus $g$ entering at
$O(N^{2-2g})$.
There is even a clear picture of what the strings are in the
field theory, because we can re-interpret the Hilbert space of states at a
fixed time in string language.
For YM$_2$ quantized on $S^1\times I$, a `string' will be a loop about the
$S^1$
with a specified winding number $n$ (call it $L(n)$), and the Hilbert space is
a Fock space of bosonic loops.
The Wilson loop operator
$W_n=\Tr \exp i\int_{L(n)} d\sigma A_\sigma(\sigma,\tau)$
is then $\alpha_{-n}+\bar\alpha_{n}$ in terms of
bosonic creation and annhilation operators.
These are defined as in two-dimensional conformal field theory,
but should not be thought of as modes of a local field in the original two
dimensions.
Although this description is implicit in section 5 of \gt,
the derivation of \refs{\douglas,\polym}, though perhaps overkill for YM$_2$,
will have a clear generalization to CSW theory, so we briefly review it.
On the cylinder and in $A_\tau=0$ gauge, YM$_2$ reduces to the group quantum
mechanics of the holonomy
$U=\exp i\int_0^{2\pi} d\sigma A_\sigma(\sigma,\tau)$.
The treatment of \BIPZ\ can be applied to this problem, producing $N$ free
non-relativistic fermions, whose large $N$ limit is the same as in section 3
but with non-compact momentum.
Since the ground state has a well-defined Fermi surface, the finite energy
excitations are those of free, quasi-relativistic fermions.
Thus standard two-dimensional bosonization can be applied
just as it was for the matrix model in \bosecone, producing the kinematic
result we already stated, and turning the Hamiltonian into the interacting
Jevicki-Sakita bosonic Hamiltonian (in compact space and with zero potential).

For everything we say below, it is important to realize that this bosonization
was explicitly defined as an expansion around a particular Fermi surface, here
$\rho_0$.
The usual treatment in terms of a field
$\phi(\theta)=\int dp~\rho(p,\theta)-\rho_0(p,\theta)$ will break down if the
amplitude of $\phi$ becomes large enough to violate the constraint
$\rho(p,\theta)\ge 0$.  A single quantum excitation has amplitude $O(1/N)$
compared to $\rho_0$ and so for $O(N^0)$ excitations this will not happen, but
in trying to construct classical solutions by perturbation theory (say around
$\rho_0$) it can happen.

Treating the interaction term (which is $O(1/N)$) using time-ordered
perturbation theory, and expanding everything in oscillators,
one can make a direct correspondance between terms in this perturbation theory
and the world-sheets of \gt, as is done in \polym.
Free propagation corresponds to free string evolution, while an interaction
will correspond to a world-sheet branch point or `tube'.
Thus the possibility of a string interpretation at all orders in $1/N$ rested
on the fact that bosonization was exact to all orders in $1/N$.
We can even test this assertion, because there are quantities in YM$_2$ for
which the string picture breaks down, namely the partition function $Z(S^2)$
for area (time in the equivalent quantum mechanics) $g^2 A<\pi^2$, the weak
coupling phase. \dk~
The bosonization around $\rho_0$ is valid for $g^2 A$ large, and underlies a
string interpretation which produces a sum over terms coming from $n$-fold
covers of the sphere and weighed by $\exp -n g^2 A$.
This sum reproduces correct expectation values as long as it is not necessary
to take into account the constraint $\rho(p,\theta)\ge 0$.
However, the weak coupling phase is governed by the saddle point
\eqn\rhoweak{\rho_w(q,p;t) = \cases{
1\qquad {\pi\over y}q^2 + \pi y p^2 \le 1\cr
0\qquad {\rm otherwise}}}
with $y=t(1-t) g^2 A/\pi$ and $t=A_1/A$ for a Wilson loop enclosing area $A_1$
(from (50) in \dk, which should have an extra $\half$ in the exponential).
This is unrelated as an analytic function to $\rho_0$ and inaccessible to
series expansions around this point.

It would be quite interesting if a bosonization could be defined around the
Fermi surface $\rho_w(q,p;t)$.  Perhaps this could underlie a `string
reformulation of the weak coupling phase of YM$_2$,' and it might be that this
would be a better prototype for the higher dimensional case.
One problem is that $\rho_w(q,p;t)$ is not a static solution of the classical
fermion theory, so it is not clear we can bosonize around it,
and there are no static solutions with the same qualitative behavior.
This problem is rather specific to this YM$_2$ calculation and would not be
present if we were working around a static ground state.

We briefly reviewed the reformulation of the YM$_2$ Hilbert space and
Hamiltonian in string language; of course there are other approaches
\refs{\gt,\cmr}\ and each illustrates interesting features of the problem.
However, we will propose a rather strong hypothesis about the importance of the
Hilbert space reformulation:
we believe that any string interpretation which could be reproduced by a local
world-sheet path integral
must allow a description of the Hilbert space
(for any choice of quantization surface) in string language,
simply because we require that time evolution be well defined in string
language.
It does not seem sensible to try to make a precise definition covering all
string theories, but essentially we mean that there exists a basis for the
Hilbert space labelled by occupation numbers associated with `loops,' which
for gauge theories should be continuous loops in space.
The correspondance must be one-to-one.
We can imagine assigning more than one string state to the same field theory
state,
but this would produce a string theory containing our original field theory and
not literally equivalent to it.  We would have to recheck the usual axioms of
quantum theory for the new theory.

For the matrix model, although it may not be the best picture for contact with
two-dimensional string theory, loops can be labelled by their world-sheet
length, and the fields can be transformed to this representation,
as was done in
\ref\ms{G. Moore and N. Seiberg, Int.J.Mod.Phys.A7 (1992) 2601.}
for $c=1$ (and as appears in
\ref\kost{
I. Kostov, Nucl.Phys.B376 (1992) 539;\Cr
I. Kostov and M. Staudacher, Phys.Lett.B305 (1993) 43.}.
It is easier for the $c=0$ loop equations.)
We then have a very similar formal relation between the exact free fermion
Hilbert space and this `string interpretation.'
A Hilbert space interpretation can also be made in critical string field
theory, most clearly in light-cone gauge, and there the `loops' are not
continuous but are the more abstract Fock basis of first quantized string
theory.  Whether this can be done in equal-time quantization is a rather deep
question.

For gauge theory, there is no difficulty in defining the state at a given time,
making our hypothesis even better motivated.
Unfortunately, the naive interpretation of this for $D>2$ Yang-Mills theory,
taking all loops $[0,2\pi]\rightarrow \BR^{D-1}$ (up to reparameterization) as
independent and building a bosonic Fock space on this, is almost certainly
wrong,
and making a precise description is a central issue in making sense of `QCD
string' in $D>2$.
Although there are well-known disadvantages to Hamiltonian treatments of
quantum field theory, at least the approach we are advocating here attempts to
confront the central issue head-on: that just because the observables are
naturally formulated as functionals of a loop, does not immediately imply that
the dynamics is naturally equivalent to a dynamics of loops.

{}From the YM$_2$ sphere problem
we see that we should break this problem into two steps.
One can first try to establish it in an expansion around $\vev{W[L]}=0$.
One then needs to verify that the same Fock space description works around the
background of interest, or modify it to work.
In particular, if a large $N$ transition separates a background of interest
$\vev{W[L]}=\phi_0$, from $\vev{W[L]}=0$ for which we know how to reformulate
the perturbation theory as string theory, analogous to the transition between
weak and strong coupling in YM$_2$, we should expect to be able to use the same
formal structure (or `classical string field theory') of loop equations and
Hamiltonian, but starting with perturbation theory using
$\vev{W[L]}=\phi_0+\delta\phi$ and looking for a modified string reformulation.
If $\phi_0$ is a ground state, this perturbation theory is surely well defined.
As for the question of whether the new perturbation theory has a string
reformulation, while it is impossible to answer on general grounds, considering
 the $c=1$ matrix model and modifications of it with general potentials might
suggest a reason to be optimistic.
There, very different vacua can be possible in the same model, and typically
the same formal technique of bosonization can be applied around each vacuum to
produce a `string theory' (a bosonic theory with the Jevicki-Sakita
Hamiltonian).
Furthermore, it should not be necessary to use the true ground state for this
purpose but just one in the same (at this point rather ill-defined) `class',
just as the YM$_2$ string interpretation was justified by bosonizing around
$\rho_0$, and allowed calculating the strong coupling result for $Z(S^2)$ at
finite $g^2 A$, for which $\rho(t)\ne \rho_0$.
The string perturbation theory we define around the reference state should sum
to the correct amplitudes.
This is important if we hope to solve a theory by using a string reformulation,
rather than the other way around.

It seems reasonable to expect
that if we can reformulate the QCD Hilbert space in terms of `QCD
strings' at short distances, since this is a kinematic problem,
locality will allow us to infer the complete answer.
Thus this could be studied with perturbative methods.
In fact, there is a stronger version of this
statement, which perhaps has not been sufficiently explored.
One can imagine a world, as described quite vividly in
\ref\georgi{H. Georgi, {\it Weak Interactions and Modern Particle Theory},
Benjamin-Cummings, 1984; section 3.2.},
in which the QCD coupling at the scale set by the light quark masses was small,
say $\alpha_s(m_q)=\alpha_{EM}$.
(Or, just imagine that the top quark is the lightest quark.)
Although it is not proven, the picture supplied by perturbative QCD seems quite
reasonable for this world.  If we believe an exact reformulation of QCD as a
string theory is possible, the string theory should describe this regime as
well, and although it is not completely clear that perturbative QCD produces
the
same picture in the large $N$ limit, understanding this limit and reformulating
it as a string should be a much more tractable problem than non-perturbative
QCD.  If it doesn't work, it seems likely that even if some sort of `string
reformulation' exists in other regimes, it will break down in many cases of
physical interest.
\newsec{An attempt at a CSW string interpretation}
{}From a mathematical point of view, a string interpretation for CSW theory
might be quite attractive.  Since we would reproduce the complete double
expansion of the topological invariants computed by CSW theory in $1/N$ and
$x=N/(N+k)$, we would have reformulated all of the information in the
two-parameter family of invariants in terms of closed string theory.

The intuitive picture of the string interpretation we are trying to construct
is rather unclear.  We clearly expect a topological theory in space-time, and
one
might think that the world-sheet path integral should localize on topologically
distinct classes of embeddings.
Now $Z(S^3)$ is quite non-trivial while $Z(S^2\times S^1)$ is trivial, and it
is not clear what topological classification of embeddings of Riemann surfaces
into three-manifolds would lead to such a result.

Lacking an intuitive picture,
we will start from the quantum free fermion theory and
the $1/N$ expansion of the observables, and try to implement the approach of
section 4.
Acting on the state $\ket{0}$ and finite excitations around it,
the standard (or `quasi-relativistic') bosonization will apply,
exactly as in \douglas.
The state $\ket{0}$ is produced by the path integral on
the solid torus; let us take the $a$ cycle as non-contractible.
We will divide the second quantized fermions into `left-movers' and
`right-movers' acting on the neighborhood of the two Fermi surfaces, and write
\eqn\bilin{W_{n,0} = \sum_m B^+_{n-m} B_m \equiv
\alpha_{-n}+\bar\alpha_n}
in terms of the standard bosonic operators.
Following \refs{\polym,\douglas} we interpret this as a sum of an operator
which creates a string winding $n$ times about the $a$ cycle and an operator
which destroys a string winding $-n$ times about the $a$ cycle.
Thus a Wilson loop winding about the $a$ cycle does exactly what we might
expect from our YM$_2$ experience.  For Wilson loops in representations with
$O(N^0)$ highest weight, the fusion algebra reduces to the usual
Littlewood-Richardson coefficients, which are reproduced in the usual way.
\douglas

Let us now consider a Wilson loop winding about the $b$ cycle.
This can also be expressed using bosonization.  Let $\theta=2\pi q$,
then we have
\eqn\bbilin{\eqalign{W_{0,n_b} &=
\sum_m e^{2\pi i n_b m/(N+k)} B^+_{-m} B_m
\qquad\qquad{\rm (non-relativistic)}\cr
&=N \int_{-x/2}^{x/2} dp~ e^{2\pi i n_b p}
\qquad\qquad\qquad\qquad{\rm (relativistic).}\cr
&~~~+\int d\theta~e^{i\pi n_b x} \psi^+(\theta+{2\pi n_b\over N+k})\psi(\theta)
+
e^{-i\pi n_b x}\bar\psi^+(\theta-{2\pi n_b\over N+k})\bar\psi(\theta) }}
Applying $\psi(\theta)=:\exp i\phi(\theta):$, etc...
gives
\eqn\bboson{\eqalign{W_{0,n_b} &=
{i N\over \pi n_b} \sin {\pi n_b x} \cr
&~+
(N+k) {e^{i\pi n_b x}\over 2\pi n_b}\int d\theta~
:\exp i\sum_{p\ge 1} {1\over p!} \left({2\pi n_b\over N+k}\right)^p
\partial^p\phi(\theta):~+ {\rm c.c.}\cr
&=
{i N\over \pi n_b} \sin {\pi n_b x} +
{2\pi in_b e^{i\pi n_b x}\over N+k} L_0 + {e^{i\pi n_b x}\over 6}\left({2\pi
n_b\over N+k}\right)^2
\int(\partial\phi)^3 + \ldots~.}}
It acts on the same Hilbert space as the $a$ Wilson loops,
namely a Fock space with basis elements $\ket{N_1,N_2,\ldots}$
labelled by the number $N_n$ of strings winding $n$ times about the $a$ cycle.
The leading term in $1/N$ is the disconnected contribution.
The leading connected term preserves the string number, while subleading terms
need not.

In one way this is entirely natural as the $b$ cycle is contractible.
A $b$ Wilson loop acting on the solid torus should act trivially, in a way
determined by its framing.
A $b$ loop acting on states produced by the action of $a$ loops can be
evaluated by
using the commutation relations until it is contractible.
In world-sheet terms this could be modeled with a direct interaction where the
$b$ Wilson loop pierces the world-sheet created by the $a$ string.

However, this makes the string interpretation very dependent on the particular
fermionic state we are working around.  For example, there is another state in
which only the $b$ cycle loops create and destroy strings, namely the
$S$-transform of this one.
The information in the state which determines which strings act non-trivially
is clearly the choice of Fermi surface.
This information is also present in our `classical string field theory'
description, in the expectation value of $w$.
In string terms we can say that `the state $\ket{0}$ contains a
condensate of $b$ loops.'  Now in conventional string field theories, there is
a field for every elementary excitation, and non-zero vacuum expectation values
are not surprising.  What is strange here is that there is no additional
quantum excitation corresponding to a $b$ loop.   In this sense a $b$ loop does
not create or destroy a string.  The first picture that comes to mind, in which
a $b$ loop bounds a disk, is not right because there is no field theory
counterpart of the string whose propagation we are postulating in this picture.

Now, in one's usual picture of the world-sheet path integral,
all Wilson loops correspond to boundaries of the world-sheet.
Combining this with the assumptions that the string
theory is free at leading order in $1/N$ and that the world-sheet is
continuously embedded in the target space implies that a Wilson loop operator
must create or destroy a string.

It seems that the most attractive way out is to say that $b$ loops do not bound
world-sheets.  Instead, they interact with world-sheets that they pierce.
If we are willing to accept a string theory which is defined as a different
perturbative expansion (in $1/N$) around each `background classical string
field', then this picture seems potentially consistent, and should be further
explored.
A major difficulty is that our formalism relies so heavily on the $2+1$
dimensional splitting between space and time, which could be done in many ways.
If a string reformulation exists, it should be possible to make this split, but
in the end a satisfactory reformulation must not depend on such a choice.
What we can say so far is that observables which can be formulated as operators
acting on $H(T^2)$ have a well-defined action on the string Hilbert space.
However, since typical Wilson loops cannot be embedded into $T^2$ without
self-intersecting, we cannot be sure that the `matrix elements' we have defined
can be reproduced in a way consistent with three-dimensional locality.
To study this question we need to reformulate observables which take (for
example) $H(T^2)\rightarrow H(T^2,R,\bar R)$.

Conceivably these results point to a new type of
statistics possible for strings in three dimensions.  If this were to exist it
might be relevant not only for gauge strings but for fundamental (and
effective) string theories as well.

Even this rather odd `string formulation' is not universally valid.
A simple example illustrating this is the partition function
on $T^3$.
This is simply the dimension of the Hilbert space $H(T^2)$:
\eqn\ztthree{Z(T^3) = {N+k \choose k} \equiv e^{F({T^3})}.}
The expansion of $F({T^3})$ in $1/N$ was found in \peri\ and begins at $O(N)$.
This is subleading in $1/N$ compared to $F(S^3)$ and
we should not expect the classical description to reproduce it.
In YM$_2$ subleading answers were reproduced by the quantized bose theory,
but this one is obviously impossible.
Even a single bosonic mode will act on an infinite-dimensional Hilbert space,
and there is no way that we can reproduce a formula like
$\Tr|_{H(T^2)} \sim \exp N$.
This argument may seem glib -- after all we are taking the limit
$N\rightarrow\infty$ in which the CSW Hilbert space is infinite-dimensional.
We believe it is correct, however.
The essential difference between the systems where bosonization is valid to all
orders in $1/N$ (for example YM$_2$) and CSW theory is that in YM$_2$, all
traces over the Hilbert space are weighed with a Boltzmann weight $\exp -\beta
H$ with an $H$ which is $O(N)$ for states with $O(N)$ excitations from the
vacuum.  Thus we can consistently take the $N\rightarrow\infty$ limit and drop
these states before reformulating the theory.  Since CSW theory has no
Hamiltonian there are observables which see all the states, and we would
require a bosonization which is simultaneously valid for states differing by
$O(N)$ excitations.  Such a bosonization is not known to exist.

Perhaps a better interpretation of $Z(T^3)$
is to think of it as an infinite temperature limit of a more conventional QFT
partition function.
It would thus be analogous to the unconfined phase in more conventional gauge
theories (we will not try to make this idea precise), and we might say that the
failure of a string interpretation here was to be expected.
The $O(N)$ behavior of the free energy (which is simply the entropy), although
completely obvious in the fermionic language, is quite unusual in a pure gauge
theory even with this interpretation.

Although we do not want to make too much of the analogy,
this model is a good illustration of the point made by a number of authors
\nref\natpol{M. Natsuume and J. Polchinski, ``Gravitational Scattering in the
$c = 1$ Matrix Model,'' hep-th/9402156.}
\nref\leemende{J. Lee and P. Mende,  ``Semiclassical Tunneling in
(1+1)-dimensional String Theory,'' Phys.Lett.B312 (1993) 433.}
\refs{\leemende,\morewadia}
and particularly \natpol~
about the $c=1$ matrix model, that the description in terms of a `string field'
depending only on $\lambda$ is not complete, first because it cannot describe
the most general Fermi surface, and second because it does not uniquely
determine effects of $O(\exp -N)$.
The interpretation of these facts is still mysterious in the $c=1$ model,
though we agree with \natpol\ that a reasonable conclusion to draw
(as has been suggested before, e.g. in
\ref\cargese{{\it Random Surfaces and Quantum Gravity}, eds. O. Alvarez, E.
Marinari and P. Windey, Plenum 1991, contributions by S. Shenker and D.
Gross.})
is that the fundamental variables of a `string theory' need not be strings.
In large $N$ CSW theory, it is necessary to describe more general Fermi
surfaces, and states differing by $O(N)$ excitations.
If we could give them a string interpretation here, perhaps this would suggest
new interpretations to consider for $c=1$.

Our present belief
is that there is no string reformulation allowing us to go from one Fermi
surface to a qualitatively different one.
As we argued in section 4, this phenomenon may be a prototype for thinking
about and dealing with large $N$ transitions in higher dimensional theories.

To summarize, although CSW theory has a good large $N$ limit, it
seems to be an interesting test of the hypothesis of section 4, in the negative
sense.  Thus it would be quite interesting to find a string interpretation
anyways (presumably refuting the hypothesis), or to make sense of the `mixed'
formulation we were left with in section 5 (in which a string interpretation
defined as a different expansion about each `background' was proposed) in a
three-dimensionally covariant way.

\medskip
We thank T. Banks, S. Shenker, E. Witten
and especially G. Moore for invaluable discussions.
%\vfill\eject
\listrefs
\end